# Enhancing and Analyzing Search performance in Unstructured Peer to Peer Networks Using Enhanced Guided search protocol (EGSP)

Anusuya.R[1], Dr.Kavitha.V[2], Mrs. Golden Julie.E[3]

[1] University Department, Anna University Tirunelveli[1]
Tirunelveli, Tamilnadu, India

[2] University Department, Anna University Tirunelveli[2]
Tirunelveli, Tamilnadu, India

[3] University Department, Anna University Tirunelveli[3]
Tirunelveli, Tamilnadu, India

**Abstract**

*Peer-to-peer (P2P) networks establish loosely coupled application-level overlays on top of the Internet to facilitate efficient sharing of resources. It can be roughly classified as either structured or unstructured networks. Without stringent constraints over the network topology, unstructured P2P networks can be constructed very efficiently and are therefore considered suitable to the Internet environment. However, the random search strategies adopted by these networks usually perform poorly with a large network size. To enhance the search performance in unstructured P2P networks through exploiting users' common interest patterns captured within a probability-theoretic framework termed the user interest model (UIM). A search protocol and a routing table updating protocol are further proposed in order to expedite the search process through self organizing the P2P network into a small world. Both theoretical and experimental analyses are conducted and demonstrated the effectiveness and efficiency of the approach.*

*Keywords: Peer to Peer Networks, Self-Organization, and User's Common Interest.*

## 1. Introduction

Peer-to-peer (P2P) networks have become, in a short period of time, one of the fastest growing and most popular Internet applications [6]. A class of applications that takes advantage of resources like storage, CPU cycles, content and even human presence available at the edges of the Internet.One fundamental challenge of Peer to Peer networks is to achieve efficient resources discovery. Those networks can be largely classified into two categories, namely, structured P2P networks based on a distributed hash table (DHT)[21] and unstructured P2P networks based on diverse random search strategies (e.g., flooding)[3]. Without imposing any stringent constraints over the network topology, unstructured P2P networks can be constructed very efficiently and have therefore attracted far more practical use in the Internet [1], [2] than the structured networks. Peers in unstructured networks are often termed blind, since they are usually incapable of determining the possibility that their neighbour peers can satisfy any resource queries. An undesirable consequence of this is that the efficiency of distributed resource discovery techniques will have to be compromised. The fundamental idea of this paper is that the statistical patterns over locally shared resources of a peer can be explored to guide the distributed resource discovery process and therefore enhance the overall resource discovery performance in unstructured peer to peer networks.

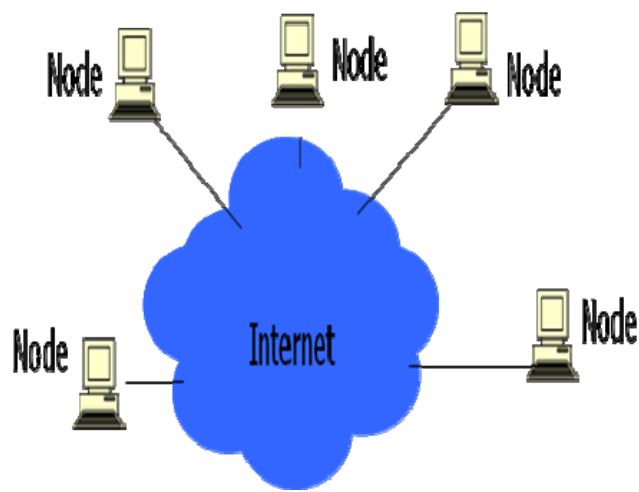

Three essential research issues have been identified and studied in this paper in order to save peers from their blindness.

The first research issue questions are the practicality of modelling users' diverse interests. To solve this problem, the user interest model (UIM) based on a general probabilistic modelling tool termed Condition Random Fields (CRFs)[14]. With UIM, we are able to estimate the probability of any peer sharing a certain resource (file) $f_j$ upon given the fact that it shares another resource (file) $f_i$. This estimation further gives rise to an interest distance between any two peers. Conditional random fields, a framework for building probabilistic models to segment and label sequence data. Conditional random fields offer a unique combination of properties: discriminatively trained models for sequence segmentation and labelling; combination of arbitrary, overlapping and agglomerative observation features from both the past and future; efficient training and decoding based on dynamic programming; and parameter estimation guaranteed to find the global optimum.

The second research issue considers the actual exploration of users' interests as embodied by UIM. For this greedy file search protocol is presented for fast resource discovery. Whenever a peer receives a query for a certain file that is not available locally, it will forward the query to one of its neighbours that have the highest probability of actually sharing that file.

The third research issue is that the search protocol alone is not sufficient to achieve high resource discovery performance. This paper proposes a routing table updating protocol to support our search protocol through self organizing the whole P2P network into a small world[11],[16],[18]. In a P2P network, queries handled by a peer may be satisfied by any peer in the network with uneven probability.

## 2. Enhanced Searching Protocol Based On UIM Model.

This section presents and analyzes the guided search solution that we proposed for resource discovery in unstructured P2P networks.

### 2.1 Peer to Peer Network

In recent years, Peer-to-Peer (P2P) technologies have become increasingly popular. A P2P system can be defined as a distributed network architecture, whereby participants share a part of their own hardware resources, such as processing power, storage capacity, or network bandwidth. The shared resources are necessary to provide the service and content offered by the network, such as file-sharing. The service or content provided by the P2P network is accessible by other peers directly, without passing intermediary entities. Peer-to-Peer (P2P) systems make it possible to harness resources such as the storage, bandwidth, and computing power of large populations of networked computers in a cost-effective manner. Actually P2P is a decentralized and distributed and here all the nodes are equivalent.

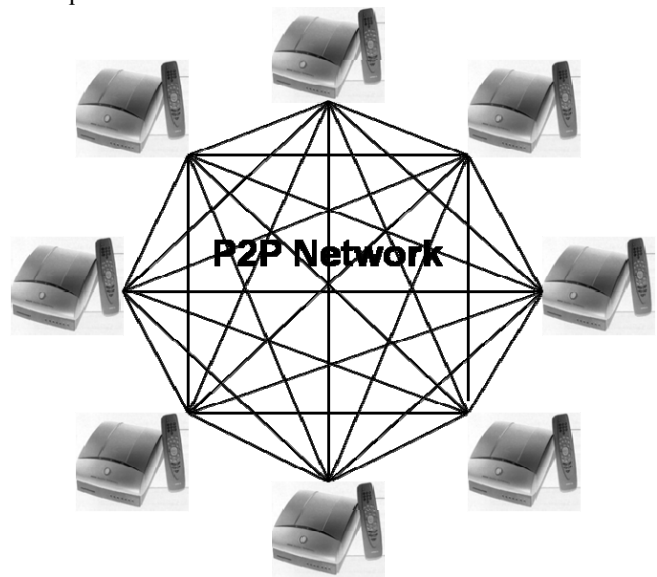

No centralized client-server scheme and network of equal "peer" nodes serving either as clients or servers to other nodes. In structured P2P systems, data items are spread across distributed computers (nodes), and the location of each item is determined in a decentralized manner using a distributed hash lookup table (DHT).Structured P2P systems based on the DHT [21] mechanism have proven to be an effective design for resource sharing on a global scale and on top of which many applications have been designed such as file sharing, distributed file systems, real-time streaming, and distributed processing.

### 2.2 User Common Interest Model (UCIM)

This section deals with one essential problem as to how user's interests can be modeled properly. Our UIM aims at characterizing users' common interest patterns within a probability-theoretic framework. It is adapted from a general probabilistic modeling tool termed Conditional Random Fields (CRFs)[14].Similar with CRF; UIM defines a log-linear conditional probability distribution Pr $(f_j\backslash f_i)$ between any two files $f_i$ and $f_j$

In this paper, Pr $(f_j\backslash f_i)$ refers to the probability that any user can be interested in sharing file $f_j$, given the fact that he/she shares another file $f_i$.Probabilistic inference in UIM is very efficient, without relying on any independence assumptions as required by other probabilistic modeling techniques such as the Hidden Markov Model (HMM)[22]. It should also be noticed that to propose a generic probabilistic model suitable for a wide range of applications is not the focus of this paper. For the purpose of this paper alone, we found that UIM is expressive enough to model users' common interests and to guide the resource discovery process. In practice, every file shared through a P2P network can be uniquely described with a list of attributes. The key structure for estimating $Pr(f_j \backslash f_i)$ in UIM is the feature function. Each feature function stands for a certain domain-specific criterion, which is essential for evaluating $Pr(f_j \backslash f_i)$. If the criterion is satisfied, F(.) will return 1. Otherwise, 0 will become the output of F(.).The definition of feature functions forms the structure core of UIM, which is domain dependent and can be constantly learned via model learning algorithms. Based on this UIM, $Pr(f_j \backslash f_i)$ is to be evaluated as

$$Pr(f_j|f_i) = \frac{1}{Z(f_i)} \cdot 2^{\sum_{h=1}^{k} \lambda_h \cdot F_h(f_j, f_i)}$$

Where $Z(f_i)$ is the normalization factor such that all probability values under the condition of file $f_i$ add up to 1. For every $F_h(f_j,f_i)$ the corresponding weight $\lambda_h$ in (1) measures the probabilistic significance of the criterion imposed by $F_h(f_j,f_i)$. These weights are to be determined through maximum likelihood training algorithms based on available training data. The details will be omitted here. UIM actually represents a log-linear probability distribution.

## 2.3 Learning Methods

The primary concern of this paper is to manage network topology and to enhance resource discovery performance with the help of UIM. However, to make our discussion complete, this section will briefly introduce the process through which UIM can be learned and updated. The basic design principles of peer to peer networks, UIMs are better to be learned locally by every peer. However, in order to ensure that these locally maintained UIMs will remain consistent with each other, designated servers will also be employed to fulfil certain computation intensive learning tasks. The establishment of UIM comprises two levels of learning tasks:
1) Structure learning to determine a group of feature functions and
2) Parameter learning to determine the weight $\lambda_h$ associated with every feature function in the group. In general, Structural learning is a challenging task that demands intensive computation resources. Based on collected information, a reference UIM will be created, with emphasis on its modeling structure (i.e., feature functions). This reference UIM is then broadcast through the P2P network to update the structure of those UIMs locally maintained by every peer. Hence the learning methods are used to manage the network topology and to enhance or to improve the performance of resource discovery.

Meanwhile, the structure of a UIM,which captures essentially the dependence relationships among a group of file attributes, is expected to change infrequently. A UIM structure update through servers will not introduce considerable communication cost. The model (UIM) based on a general probabilistic modelling tool termed Condition Random Fields (CRFs).In comparison with structural learning, parameter learning usually happens more frequently UIM serves essentially as a measure of the distance between any two peers or any two files. In cooperation with a proper strategy for updating routing tables, a small-world network that guarantees search efficiency can be formed and 3)The Enhanced Guided Search Protocol. In this section, a file search protocol is presented to regulate the activities of every peer p in a P2P network upon receiving a query $q = <p;f; h_q;TTL;t_s; t_e>$.

**Protocol for Enhanced Searching:**

Require
Step (1) Query $q=<p_q,f_q, p_q ,TTL,t_s,t_e>$received by a peer p.
Step (2) The routing table $R_p$ of peer p.
Step (3) UIM
1. Add peer p to the search history $h_q$.
2. If file $f_q$ is locally stored in peer p, Inform peer $p_q$ that the search is successful.
3. Else if the size of $h_q$ is greater than TTL, Inform peer p that the search fails.
Step (4) Else
  a. Order all the routing entries $E_p=<p_e ,f_e>$ of $R_p$
    decreasingly based on $pr(f_q|f_e)$.
  b. Iterate over every routing entry $E_p=<p_e,f_e>$,starting
    from the one with the highest $pr(f_q|f_e)$.
    1. If $p_e$ has never been visited before according to
    $h_q$,forward the query q on to peer $p_e$.
    2. Else continue iteration with next routing entry.
  c. If no entry is chosen at step 4.b, forward query q on to
    $p_e$ with highest $pr(f_q|f_e)$.

The local decision involved in the search protocol demands three main types of information:
1)The search history $h_q$ stored in the query q,
2) The routing table $R_p$ of the peer p that handles the query q, and
3) The UIM $h_q$ and $R_p$, which are readily available in many peer to peer networks.

## 3. Proposed work

### 3.1 The Updating Routing Table Protocol (URTP)

In this section, a protocol for updating routing tables will be presented and analyzed. An uneven updating problem will also be highlighted, and a filtering mechanism will be further introduced to tackle this problem. This paper considers a loosely connected peer to peer network. We use p to denote a single peer in the network. P is further utilized to denote the set of all peers in the network. The main type of resource, namely, a data file, is represented by f. For every peer p, $F_p$ is used to represent the group of files shared by p. In order to conduct distributed search over the P2P network, every peer p maintains locally a list of neighbor peers. This list serves as the routing table for peer p, denoted by $R_p$. There is an upper bound Br on the size of any routing table $R_p$, while the size is measured in terms of the number of entries in $R_p$. An entry $E_p$ of $R_p$ is a tuple of two elements: < p0; f0 >. It represents a link from peer p to another peer p0 that shares file f0. In order to locate (or discover) any file under request, the user of a peer p, denoted by up, sends out a query to the network. A query that originated in peer p is represented by $p_q$ and is a tuple of six elements:< $p_q$ ; $f_q$; $h_q$;TTL;$t_s$;$t_e$ >. Here, p stands for the peer that issued the query $q_p$. f is the file requested by the query. $h_q$ records the search history, which is a list of peers that have processed the query previously, including peer p itself. In order to prevent a query from incurring too much traffic in the network, time-to- live (TTL) in a query defines an upper bound on the allowable size of $h_q$, $t_s$ refers to the time when the query is issued, while $t_e$ is the time when the query is completed. A query is completed successfully if the requested file f has been identified. On the contrary, the query is failed if the size of $h_q$ exceeds the TTL.

Upon receiving a query q, a peer p needs to perform several basic operations: 1) append itself to the search history $h_q$, 2) search the requested file f among its locally shared files (i.e., local repository), and 3) forward the query to one of its neighbor peers. Each forwarding operation is termed a hop. At the time when query q is finished, the number of hops NOP becomes an important measure of the search performance. In practice, we hope that NOP for average search tasks could be as low as possible, which essentially implies that only a small group of peers, will be involved in processing any query. To summarize, there are two widely used performance metrics for resource discovery in P2P networks: NOP and search success rate. Search success rate refers to the proportion of queries that have been successful among all the queries issued by network users.

**Protocol for Updating Routing Table**

Require:(1)The search driven by query q=<$p_q$,$f_q$,$h_q$,TTL> has been successful. Peer $p_i$, which shares file $f_q$ ,has been identified.
Step 1.Repeat for every peer p in the search history $h_q$ ,
Step 2.Add a new entry $E_p$ =<$p_i$,$f_q$>to the routing table R of peer p.
Step 3.While the size of $R_p$ is greater than $B_r$.
 a. With respect to each entry $E_p$=<$p_e$ ,$f_e$>,calculate the interest distance $d(p_e. p)$.
 b. Select an entry $E_p$ =<$p_e$ ,$f_e$ > with probability proportional to $d(p_e,p)^r$.
 c. Remove $E_p$ from the routing table $R_p$

The details of our protocol for updating routing tables are described. Whenever the search process driven by any query q =< $p_q$;$f_q$; $h_q$;TTL;$t_s$; $t_e$ > is completed successfully, a new routing entry $E_p$ = < $p_i$ ; $f_q$ > , indicating that peer pi shares the queried file $f_q$, will be temporarily added into the routing table $R_p$ of every peer p recorded in the search history $h_q$. If $R_p$ is not full, no entries of $R_p$ will be removed. Otherwise, the size of $R_p$ will be reduced to below Br by deleting one or more selected entries. For our approach, with respect to each routing entry $E_p$ =< p"; f" >maintained by peer p, the interest distance between p" and p is evaluated. The probability of removing any entry is proportional to $d(p',p)^r$. Different from this approach, three competing strategies to be analyzed in this paper for updating routing tables are summarized as follows:

1. The LRU strategy. The routing entry that is least recently used to forward queries will be dropped.
2. The ECCR scheme. With a certain probability Pre, the least recently used routing entry will be dropped. Otherwise, the neighbour peer p' which has the longest interest distance from peer p, will be removed from $R_p$.
3. The distance-centric (DC) strategy. Either the peer p",which has the longest interest distance from peer p'or another peer p", which has the second longest distance, will be removed from $R_p$ of peer p,depending on a probability Prd.To make our analysis achievable, the routing table updating process will be represented through a DLM

### 3.2 Our Filtering mechanism

Specifically, network users might only query for those files directly related to their local interests. In other words, the peer p0 that is able to satisfy any query from another peer p normally is close in distance to p. Hence, newly added peers for updating a routing table are not evenly distributed across the full distance range. This uneven distribution will possibly render our routing table updating protocol ineffective. We call this problem uneven updating problem. In general, it is hard to appropriately model the probabilistic distribution over newly added neighbour peers, which might also change across time. Instead of modifying our protocol in a filtering mechanism will be presented in this section to further control the routing table updating process.The purpose of the filtering mechanism is to enforce so that our protocol can still remain effective. This is achieved by controlling which peer p'can be accepted to update the routing table of another peer p (i.e., add p' to $R_p$ of peer p). The control decision will be made based on the density of previously accepted peers around the interest distance $d(p',p)$. If this density is high, the probability of accepting p' will be lower. Conversely, if the density is low, peer p' will be accepted with a higher probability. Through this mechanism, a uniform distribution over peers accepted to update the routing table is encouraged across a wide distance range.

## 4. Performance Evaluation

Comparing with Guided Search, Routing protocol, filtering with routing updating table provides optimum results for the search performance. Initially when the queries are minimum, guided search performance was good. When the queries are getting increased ,filtering mechanism with routing updating table is the suitable one which gives the best results up to 90%.Hence it improve the searching performance of the peer. Routing updating table protocol contains the past successful search results and it is used for future references. Updating process can be taken place in each and every second.

## 5. Simulation Model

Simulation is based on NS-2 and Tcl with C++.Network Simulators such as NS-2 has been used for testing p2p protocols, while other network simulators ,like OMNeT++ have been forced to produce a simulator specifically designed for P2P systems namely oversim.We have taken X-axis parameter as queries and Y-axis as success rate. By varying different methods like guided search, simple routing and routing with filtering towards search performance with varied queries (50,100,150………)

## 6. Simulation Results

Experiments were run using different parameter, protocols and system settings. The performance analysis presented here is designed to compare the effects of different filtering mechanisms parameters such as NOP, success rate, queries etc together with P2P protocols for the improvement of search performance. In this section 6.1(a) and 6.1(b) clearly shows the optimum results.

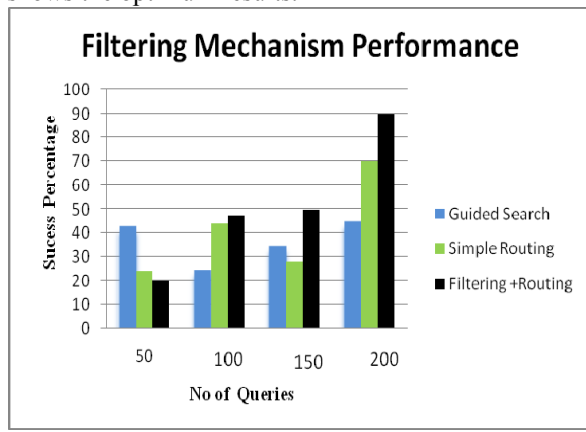

Fig 6.1(a) Performance of filtering mechanism

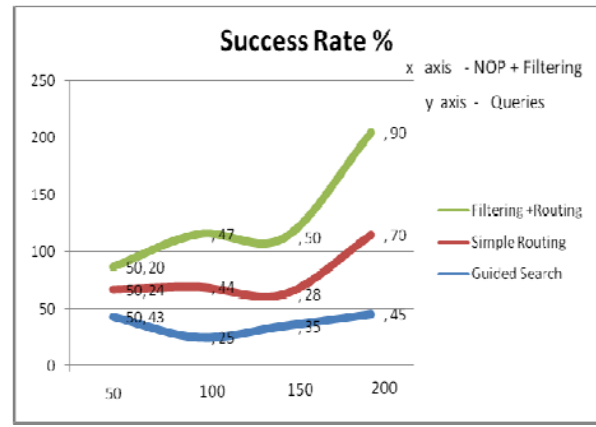

Fig 6.1(b)Success Rate performance(%)

## 7. Conclusion

Peer-to-peer networks are autonomously created, self-organizing, decentralized systems that appeal to everyday home computer users. We have shown that these networks can be organized into interest-based communities using simple formation and discovery algorithms. The search performance in unstructured P2P networks can be effectively improved through exploiting the statistical patterns over users' common interests. Specifically, the search protocol was shown to be quite efficient in small-world networks. Succeeding analysis further justifies that by using our routing table updating protocol, the P2P network will self organize into a small world that guarantees search efficiency. Common interests seek to enhance the search performance in unstructured P2P networks.

- ❖ Through exploiting users' common interest patterns captured within a probability-theoretic framework termed the user interest model (UIM).
- ❖ A search protocol and a routing table updating protocol are further proposed in order to expedite the search process through self organizing the P2P network into a small world.

Conditional random fields offer a unique combination of properties: discriminatively trained models for sequence segmentation and labeling; combination of arbitrary, overlapping and agglomerative observation features from both the past and future; efficient training and decoding based on dynamic programming; and parameter estimation guaranteed to find the global optimum.

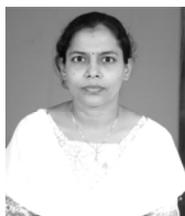

Ms.R.Anusuya received his B.E. degree in Computer science Engg in 2006 from Anna University Chennai and M.E. degree in Computer Science and Engineering in 2010 from Anna University Tirunelveli, Tirunelveli, India. Her areas of interest are Network Security, Mobile computing, Computer Networks and software Engineering. She has presented many papers in national and International Conferences in various fields. As part of this paper, she is working on developing Routing protocols for wired networks—protocols optimized for wired and that can support the shortest path and searching performance. She is a member of ISTE.

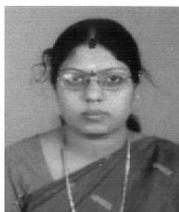

Dr.V.Kavitha obtained her B.E degree in Computer Science and Engg in 1996 from MS University and ME degree in Computer Science and Engineering in 2000 from Madurai Kama Raj University. She is the University Rank Holder in UG and Gold Medalist in PG.She received PhD degree in computer science and Engg from Anna University Chennai in 2009. Right from 1996 she is in the Department of Computer Science & Engg under various designations. Presently she is working as Asst. Prof in the Department of CSE at Anna University Tirunelveli.In addition she is the Director In-Charge of University V.O.C College of Engineering. Tuticorin.Currently, under her guidance ten Research Scholars are pursuing PhD as full time and part time. Her research interests are Wireless networks Mobile Computing, Network Security, Wireless Sensor Networks, Image Processing, Cloud Computing .She has published many papers in national and International journal in areas such as Network security, Mobile Computing, wireless network security, and Cloud Computing. She is a life time member of ISTE.

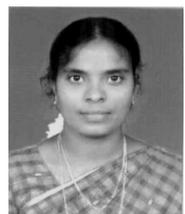

Mrs.E.Golden Julie received her B.E degree in Computer Science and Engg in 2005 from Madurai Kama Raj University and ME degree in Computer Science and Engineering in 2008 from Anna University Chennai. Currently she is Pursuing her PhD from Anna University Tirunelveli.She has published many papers in various fields. Her research area includes Data Mining, Grid Computing, Mobile Computing, Wireless Networks and Image Processing. She is a member of ISTE.